\documentclass[12pt]{iopart}
\newcommand{\eg}{{\it e.g.}}
\newcommand{\ie}{{\it i.e.}}
\newcommand{\etc}{{\it etc.}}
\newcommand{\f}{\frac}
\newcommand{\bea}{\begin{eqnarray}}
\newcommand{\eea}{\end{eqnarray}}
\usepackage{graphicx}
\usepackage{amssymb}

\begin{document}
\title[Weighted network modules]
{Weighted network modules}
\author{\it 
Ill\'es J. Farkas$^{\,1,2}$,
D\'aniel \'Abel$^{\,1}$,
Gergely Palla$^{\,1,2}$ and
Tam\'as Vicsek$^{\,1,2}$
}
$^{1}$ Dept. of Biological Physics, E\"otv\"os University
and\\
$^{2}$ Stat. and Biol. Phys. Group of HAS, 
P\'azm\'any P. s. 1A, Budapest, H-1117 Hungary
\ead{vicsek@angel.elte.hu}
\begin{abstract}
The inclusion of link weights into the analysis
of network properties allows a deeper insight
into the (often overlapping) modular structure
of real-world webs.
We introduce a clustering algorithm
(CPMw, Clique Percolation Method with weights)
for weighted networks
based on the concept of percolating $k$-cliques with
high enough intensity.
The algorithm allows overlaps between the modules.
First, we give detailed analytical and numerical results
about the critical point of weighted $k$-clique percolation
on (weighted) Erd\H os-R\'enyi graphs.
Then,
for a scientist collaboration web and a stock correlation graph
we compute three-link weight
correlations and with the CPMw the weighted modules.
After reshuffling link weights in both networks
and computing the same quantities for the randomised control graphs
as well, we show that groups of $3$ or more
strong links prefer to cluster together in both original graphs.
\end{abstract}
\maketitle

\section{Introduction}
\label{sec:intro}

Networks provide a ubiquitous
mathematical framework for the analysis
of natural and man-made systems
\cite{watts98nature,albert02rmp,mendes02advphys,newman03siam,boccaletti06physrep}. 
They allow one to picture, model and understand in a simple
and rather intuitive way the
high diversity of phenomena ranging 
from technological webs
\cite{pastor-satorras04book}
to living cells
\cite{barabasi04nrg}, 
ecological interactions
\cite{sole01ecol}
and to our societies
\cite{wasserman94book}.
The key to the applicability of
the network approach is one's ability to dissect the
phenomenon under analysis into a list of 
meaningful interacting units connected
by pairwise connections.

Over the past decade several fields
of science have been reshaped
by a flood of strongly structured experimental information.
Due to this transition, algorithms
extracting compact, informative statements from 
measured data receive a steadily increasing attention:
among such techniques the clustering of data points has become a
widely used one \cite{everitt01clustering}.
In networks clustering methods
locate network modules
\cite{girvan02pnas} 
(also called clusters or communities),
\ie, internally densely linked groups of nodes,
and lead the observer intuitively 
to a transformation
replacing the original network by its modules.
The resulting web of modules
contains ``supernodes'' (the modules)
and a link between two supernodes,
if the corresponding modules of the original network 
are linked \cite{girvan02pnas}
or overlap \cite{palla05nature}.
Interestingly, this mapping resembles a renormalisation step
from statistical physics \cite{song05nature}.
Recent practical applications
of network clustering techniques include 
the grouping of titles in a web of co-purchased books
(each cluster represents a topic)
\cite{clauset04pre},
the description of cancer-related protein
modules in a web of protein-protein interactions
\cite{jonsson06bmc}
and in stock correlation graphs
the identification of business sectors 
or the analysis of links between different sectors
\cite{onnela04epjb,kim05pre}.

A major success of the network approach to the analysis of
large complex systems has been its ability to pinpoint key
local and global characteristics based on not more than 
the bare list of interactions. This list is a ``plain'' graph, \ie,
it describes
nodes and links without any additional properties,
and has been often referred to as the
topology of interactions 
or the static backbone of the underlying complex system.
The most pronounced and widely observed 
static features are the small-world property \cite{watts98nature},
the scale-free degree distribution \cite{barabasi99science}
and overrepresented small subgraphs
(motifs) \cite{milo02science}.
In addition, correlations between neighbouring
degrees were found to
define distinct types of real-world webs
\cite{maslov02science,colizza06natphys}.
However, several important aspects
of the investigated systems
can be described only by incorporating
additional measurables, \eg, 
link weights \cite{barrat04pnas,barrat04prl,newman04pre,serrano06pre},
link directions \cite{yu06pnas,bernhardsson06pre} or 
node fitness \cite{bianconi01prl,fortunato06prl} into the models. 
Examples for the use of these characteristics are
large-scale tomographic measurements of the Internet
identifying heavily congested sections together with possible
alternative routes \cite{claffy99nature}
and the decomposition of multi-million social webs
into groups of individuals with common
activity patterns \cite{palla07phone}.

The additional graph property often providing the deepest insight  
into the dynamical behaviour is the weight of links. In the
Internet and transportation webs link weights describe traffic
\cite{pastor-satorras04book,barrat04pnas},
in social systems they represent the frequency and 
intensity of interactions
\cite{wasserman94book,newman01pre1,newman01pre2}
and in metabolic networks they encode fluxes
\cite{almaas04nature}.
Generalisations of several graph properties 
to the weighted case have revealed that, \eg,
in air transportation webs
strong links tend to connect pairs of hubs,
while in scientific collaboration graphs
the degree of a node (number of co-workers)
has almost no influence
on the average weight of the node's connections
(co-operation intensities)
\cite{barrat04pnas}.
In Ref.\,\cite{onnela05pre} 
motifs
were generalised to the weighted case
using the geometric mean of a subgraph's link weights.
With this definition the total intensity of triangles,
\ie, a generalised clustering coefficient,
was successfully applied for a weighted net of NYSE stock correlations
to find the structural characteristics and precise time
of a major crash.
Global modelling approaches to weighted graphs
include a weight-driven preferential attachment growth
rule \cite{barrat04prl} and the embedding of nodes into 
Euclidean space \cite{mukherjee06pre}.
As for weighted correlation functions,
in empirical networks they often 
depend both on the unweighted link structure (the backbone)
and the distribution of weights on
these links. Maximally random weighted networks \cite{serrano06pre}
provide a null model to separate these two effects.

As a step towards the characterisation of the
modules of complex networks,
we introduce in this paper a clustering
algorithm locating overlapping modules in weighted graphs (nodes
connected with weighted links). 
This technique, that we call the CPMw,
extends the (unweighted) Clique
Percolation Method (CPM) \cite{adamcsek06bioinf}
by applying the concept of subgraph intensity 
\cite{onnela05pre}
to $k$-cliques (fully connected subgraphs on $k$ nodes).
Similarly to the CPM, by definition the CPMw permits overlaps
between the modules, a property increasingly 
recognised in several types of complex networks 
\cite{luscombe04nature,wuchty05proteomics,pollner06epl}. 
To illustrate the use of the CPMw, 
we compute the weighted modules of two empirical networks
and investigate the correlation properties of their link weights.
Also, we provide detailed analytical and numerical results for
the percolation of $k$-cliques with intensities above a fixed
 threshold, $I$, in the weighted Erd\H os-R\'enyi (ER) graph.

\section{Definitions}
\label{sec_def}

\subsection{Local properties and correlations}
\label{subsec_basic}
Probably, the most basic properties of a node ($i$)
in a weighted network are
 its degree, $d_i$ (number of neighbours), and its strength, $s_i$ 
 (sum of link weights).
In several real systems node degrees (or strengths) are
 correlated: the network is assortative, if adjacent nodes
 have similar degrees and it is disassortative,
if adjacent nodes have dissimilar degrees.
The correlation between link weights can be studied in a very similar way.
 Two links are adjacent, if they have one end node in common,
 and link weights are assortative (disassortative) in a network, 
if the weights of neighbouring links are correlated (anti-correlated).
Moving from pairs of links to triangles, one can quantify 
the assortativity of link weights
in triangles (with nodes $i$, $j$ and $k$) by measuring 
the weight of a link, $w_{i,j}$, as a function of the geometric mean of the 
other two links' weights, $w_{i,k}$ and $w_{j,k}$:
\bea
w_{i,j} = F\bigg( [ w_{i,k} w_{j,k} ]^{1/2} \bigg) \, .
\eea
If the link weights in a triangle are similar
(or very different), then $F$ is
 an increasing (or decreasing) function.
This definition is closely 
related to the intensity, $I(g)$, of a subgraph, $g$, 
defined as the geometric mean
of its link weights \cite{onnela05pre}.

\subsection{Clique Percolation Method (CPM)}
\label{subsec_cpm}

\begin{figure}[t!]
\centerline{\includegraphics[angle=0,width=0.9\columnwidth]{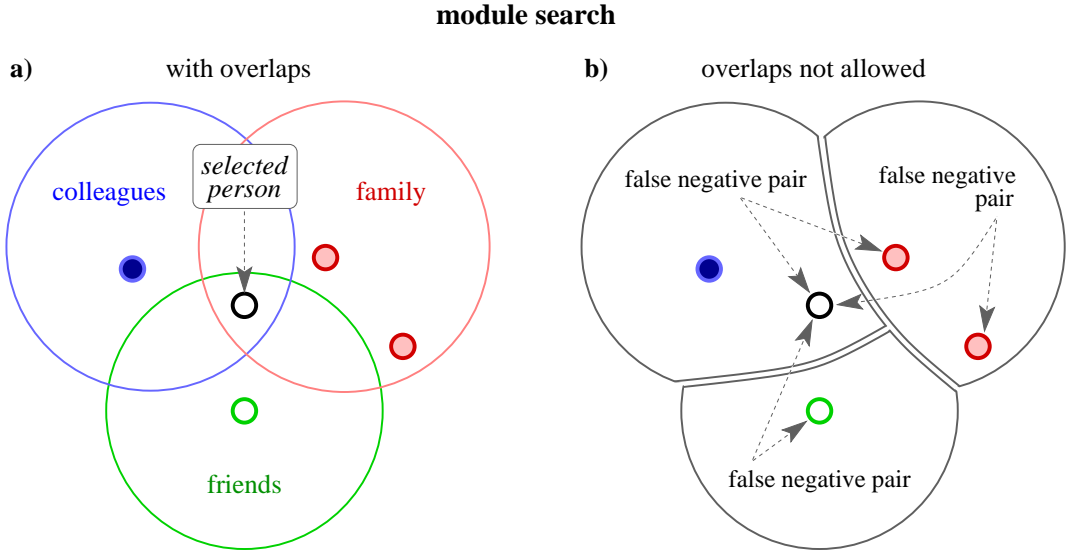}}
\caption[]{
Schematic illustration of the difference 
between module search methods.
Divisive module search techniques do not allow a
node to belong to more than one group,
which can produce a classification with high numbers of
false negative pairs.
Algorithms allowing overlaps between the modules can
significantly reduce this problem.
{\bf (a)}
Example for the overlapping 
social groups of a selected person.
{\bf (b)}
Network modules around the same person
as identified by several divisive clustering techniques.
Observe the occurrence of false negative pairs.
}
\label{fig_overlap}
\end{figure}

In many complex networks
internally densely connected groups of nodes
(also called modules, clusters or communities)
overlap.
The importance of module overlaps is illustrated
in Fig.\,\ref{fig_overlap}.
A recently introduced, link density-based module finding
technique allowing module overlaps is the
Clique Percolation Method \cite{derenyi05prl}.

The strongest possible coupling of $k$ nodes with unweighted
links is a {\it $k$-clique}:
the $k(k-1)/2$ possible pairs are all connected.
However, natural and social systems are inherently noisy,
thus, when detecting network modules,
one should not require that all pairs be linked.
In any $k$-clique a few missing links should be allowed.
Removing $1$ link from a $(k+1)$-clique leads to
two $k$-cliques sharing $(k-1)$ nodes,
called two {\it adjacent $k$-cliques}.
Motivated by this observation,
one can define a {\it $k$-clique percolation cluster}
as a maximal set of 
$k$-cliques fully explorable
by a walk stepping from one $k$-clique to
an adjacent one.
In the CPM modules are equivalent to
$k$-clique percolation clusters
and overlaps between the modules are allowed by definition
(one node can participate in
several $k$-clique percolation clusters).

With the help of the CPM, one can define 
in a natural way the web of modules as well.
In this web, the nodes represent modules
and two nodes are linked, if the corresponding modules
overlap.
In addition, the CPM has been successfully applied to,
\eg, tracing the
evolution of a social net with over $4$ million users
\cite{palla07phone} and
for highlighting which proteins -- beyond the already
characterised ones -- are possibly
involved in the development
of certain types of cancer
\cite{jonsson06bmc}.


\subsection{The Clique Percolation Method in weighted networks (CPMw)}
\label{subsec_cpmw}

The search method described in the previous section is applicable 
to binary graphs only (a link either exists or not).
Therefore, in weighted networks the
 CPM has been used to search for modules by removing links weaker 
than a fixed weight threshold, $W$, and considering the remaining connections 
as unweighted. Here we introduce an extension of CPM that takes into
 account the link weights in a more delicate way by incorporating
 the subgraph intensity defined in Ref.\,\cite{onnela05pre} into the
 search algorithm. As mentioned in Sec.\,\ref{subsec_basic}., the 
 intensity of a subgraph is equal to the geometric mean of its link weights.
 In the CPMw approach we include
a $k$-clique into a module only, if it has an intensity 
larger than a fixed threshold value, $I$.
A $k$-clique, ${\mathcal C}$,
has $k(k-1)/2$ links among its nodes ($i$, $j$)
and its intensity can be written as
\bea
I({\mathcal C}) = \
\left( \
\prod_{i<j\atop i,j\in {\mathcal C}} w_{ij} \
\right)^{\frac{2}{k(k-1)}}.
\label{eq:I}
\eea
Note that this definition is conceptually
different from using a simple
link weight threshold and then the original CPM.
Most importantly, here we allow $k$-cliques to contain links weaker than $I$
as well. 

The $k$-clique adjacency in the CPMw is defined exactly the same as in
 the CPM: two $k$-cliques are adjacent if they share $k-1$ nodes. Finally,
 a weighted network 
module is equivalent to a maximal set of $k$-cliques, with intensities
higher than $I$, that can be reached from each other via 
series of $k$-clique adjacency connections.

\subsection{Comparing the CPM and the CPMw}
\label{subsec_diff}

\begin{figure}[t!]
\centerline{\includegraphics[angle=0,width=0.98\columnwidth]{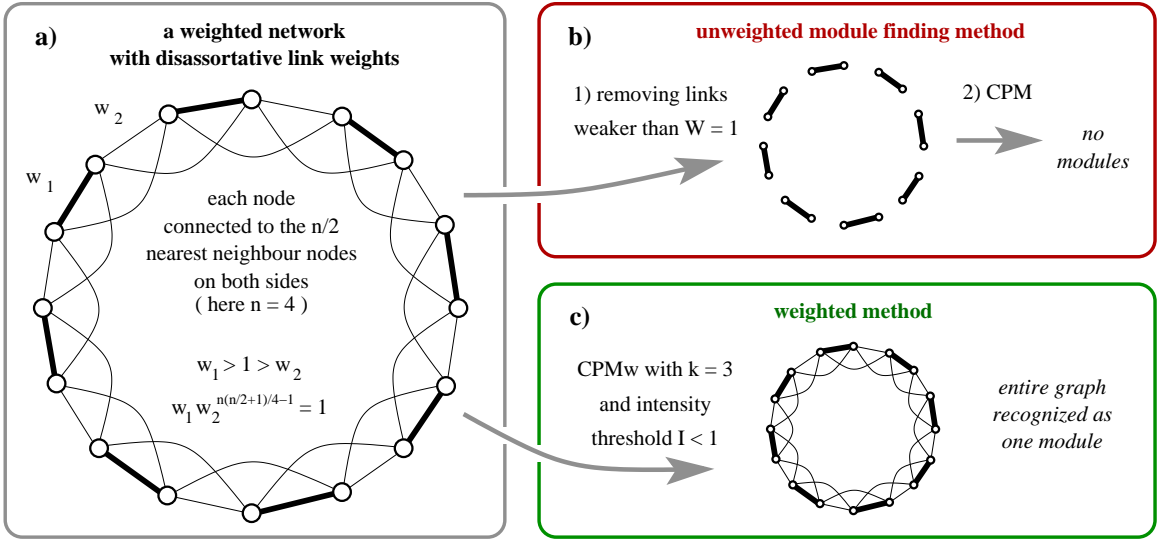}}
\caption[]{
In weighted networks with disassortative link weights,
\ie, where strong links tend to have weak links as neighbours,
the results of unweighted and weighted module finding can differ strongly.
{\bf (a)} 
Sample network with equal node degrees, $d=n$, and node strengths,
$s=w_1+(n-1)w_2$.
Each strong connection ($w_1$) has only weak ($w_2$) links as neighbours.
{\bf (b)} 
The unweighted module finding method consists of two steps
and finds no modules in the example network.
1) Links weaker than the selected threshold, $W=1$ in this case,
are deleted.
2) Applying the (unweighted) Clique Percolation Method to the
remaining links.
{\bf (c)} 
The CPMw keeps all links and finds one module containing all
nodes of the sample graph.
}
\label{fig_cpmdiff}
\end{figure}

The most important difference between the CPM and CPMw is that
 all links included in a CPM module must have weights higher than 
the link weight threshold $W$.
However, the modules obtained by the CPMw
often contain links weaker than the intensity threshold, $I$, too.
 In a weighted network where strong links prefer to be neighbours,
the above two algorithms provide similar results.
Note, however,
that the edges discarded by the first method (weight cut + CPM)
are often registered (measured) to be weaker
than $W$ only because of the inherently high noise level
of the investigated complex system. 
In comparison, the CPMw with an intensity threshold $I=W$
is more permitting and produces modules
with ``smoother'' contours.
It expands slightly 
the modules located by the CPM and
may attach to each module
additional $k$-cliques
containing weaker links.

Results from the CPM and the CPMw differ strongly for graphs
where strong links prefer to have weak links as neighbours,
\ie, links are disassortative with respect to their weights.
The assortativity of neighbouring node degrees (or strengths) 
and that of adjacent link weights are conceptually different
measures in a network.
For example, consider a circular path with an even number of
nodes and alternating
$w_1$, $w_2$ link weights
($w_1>1>w_2$; $w_1w_2^{n(n/2+1)/4-1}=1$; $n = 4, 6, \dots$)
and add weaker ($w_2$) connections between
$2$nd, $3$rd, $\dots$, $(n/2)$th neighbour nodes
(see Fig.\,\ref{fig_cpmdiff}).
In this graph node degrees and node strengths
are neither assortative nor disassortative.
Each node has a degree $d=n$ and a strength 
$s=w_1+(n-1)w_2$.
However, the strong edges ($w_1$)
have exclusively weak ($w_2$) neighbours, therefore,
link weights are clearly disassortative.
With clique size and intensity
threshold parameters $k=n$ and $I<1$ the CPMw 
recognises the entire graph as one weighted module
(Fig.\,\ref{fig_cpmdiff}c).
The corresponding unweighted search finds no modules:
If all links with weights below the
link weight threshold $W=1$ are removed,
then the remaining links will be all isolated
and the CPM finds no modules
(Fig.\,\ref{fig_cpmdiff}b).

\subsection{Further module-related definitions}
\label{subsec_further}

\begin{figure}[t!]
\centerline{\includegraphics[angle=0,width=0.9\columnwidth]{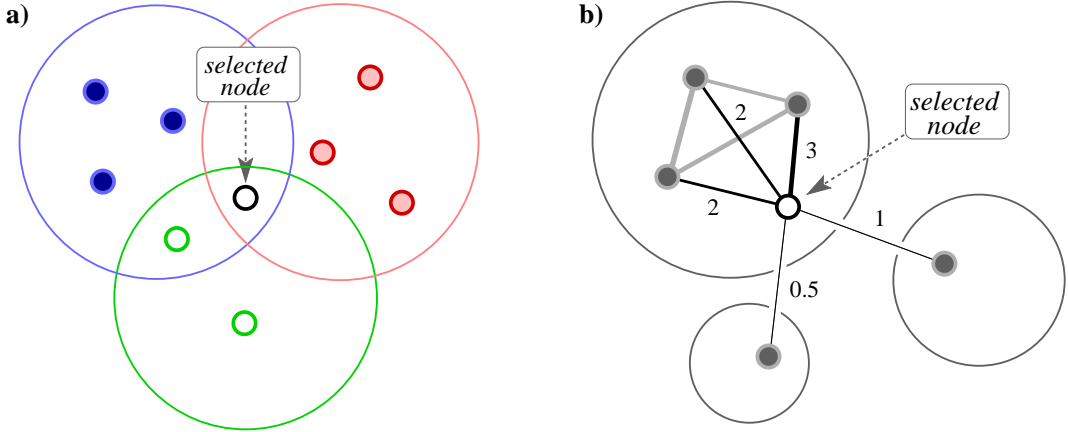}}
\caption[]{
Schematic illustrations of further
module-related quantities.
{\bf (a)}
The selected node
participates in $3$ modules, \ie,
its module membership number is $m_i=3$.
The total number of its module neighbour nodes is $t_i=8$.
{\bf (b)}
The sum of link weights (strength) connecting the selected node to
its module neighbours is $s_{i,in}=7$ and the total weight
of links connecting it to other modules' nodes is $s_{i,out}=1.5$.
}
\label{fig_defs}
\end{figure}

The number of modules that the $i$th node is contained by
is called the node's module membership number ($m_i$)
\cite{palla05nature}.
We define here the
{\it module neighbours} of the $i$th node
as the set of nodes contained 
by at least one of the modules of that node
and we will denote the number of module neighbours by $t_i$.
The total weight of links (strength) connecting the
$i$th node to module neighbours is $s_{i,in}$
and the total weight of links connecting the same vertex
to nodes in other modules is $s_{i,out}$.
See Fig.\,\ref{fig_defs} for illustrations.

\subsection{Selecting the parameters of the CPMw in real-world graphs}
\label{subsec_sel}

The CPMw has two parameters: 
$k$ (clique size) and $I$ (intensity threshold).
The optimal choice of $k$ and $I$ is the one
with which the CPMw detects the richest structure of 
weighted modules.
Here we discuss this condition from the statistical physics point of
view.

Consider a fixed $k$-clique size parameter, $k$, and a weighted graph 
with link 
weights $w_1\ge w_2 \ge \dots \ge w_L$. If $I>w_1$, then the intensity of
each $k$-clique is below the threshold, therefore no 
weighted modules are found. If, however, $I<w_L$, then any
 $k$-clique fulfils the condition for the intensity in the CPMw.
In this case often one can observe a very large weighted module
(a giant cluster) 
spreading over the major part of the network.
The emergence of this giant module (when lowering 
 $I$ below a certain critical value)
is analogous to a percolation transition.
The optimal value of $I$ is just above the critical point: 
on the one hand, the threshold is low enough to permit a huge number of
 $k$-cliques to participate in the modules, resulting in a rich module 
structure. On the other hand, we prohibit the emergence of a giant module
 that would smear out the details of smaller modules. At the critical point
 the size distribution of the modules, $p(n_{\alpha})$ is broad, usually
 taking the form of a power-law, 
analogously to the distribution of cluster sizes at the transition 
point in the classical edge percolation problem on a lattice. 

When $I$ is below the critical point, the size of the largest module, 
$n_1$, ``brakes away'' from the rest of the size-distribution
and becomes a dominant peak
far from the rest of distribution $p(n_{\alpha})$.
This effect allows one to 
determine the optimal $I$ parameter in a rather simple way.
One should start with
the highest meaningful  value of $I=w_1$
and then lower $I$ until the ratio of 
the two largest module sizes, $n_1/n_2$ reaches $2$.
However, for small 
networks this ratio can have strong fluctuations, therefore, in such cases 
it is preferable to  determine the transition point by using 
$\chi=\sum_{n_\alpha\not= n_{\rm max}} n_\alpha^2 / (\sum_\beta n_\beta)^2$, 
which is similar to percolation susceptibility.
To find the weighted modules of real-world graphs (Sec.\,\ref{sec_real}),
we first identified separately 
for each fixed $k$ the optimal $I$ value
and then we selected the $k$ parameter with the broadest $p(n_\alpha)$ 
distribution at its optimal $I$.

\section{Percolation threshold
of weighted $k$-cliques in Erd\H os-R\'enyi graphs}
\label{sec_res_ER}

An (unweighted) Erd\H os-R\'enyi (ER) graph with $N$ nodes has 
$N(N-1)/2$ possible links, each filled independently
with probability $p$.
To obtain a weighted ER graph, we assign to each link $(i,j)$
a weight, $w_{ij}$, picked
independently and randomly from a uniform distribution
on the interval $(0,1]$.
Similarly to the previous section,
we denote by $I$ the intensity threshold.
At a fixed $I$, the critical link probability, $p_C(I)$, 
of $k$-clique percolation
is the link probability where 
a giant module (containing $k$-cliques fulfilling the intensity condition)
 emerges.
A special case is $I=0$,
\ie, $k$-clique percolation on ER graphs without weights,
 for which the critical link probability can be written as
\cite{derenyi05prl}
\bea
p_C(I=0) = [(k-1)N]^{-1/(k-1)} \, .
\label{eq:pc0}
\eea

\subsection{Analytical results}
\label{subsec_an}

Below we show three analytical approximations for the critical point
of clique percolation at $I>0$.
The first is an upper bound obtained by link removal,
while the second and third are (cluster) mean-field methods.

\subsubsection{Upper bound by link removal.}

Consider a weighted ER graph, ${\mathcal G}$, with link weights as above
and remove all of its links weaker than $I$. The edges of the truncated 
weighted graph, ${\mathcal G}^{*}$, form an unweighted ER network
with link probability $p^{*}=p(1-I)$. As already noted, the intensity 
of a $k$-clique can exceed $I$ even when it contains links that are
weaker than $I$. This link removal step discards
a finite portion of the $k$-cliques ${\mathcal C}$ having $I_{\mathcal C}>I$
 from the giant (percolating) cluster of ${\mathcal G}$,
and changes the percolation threshold to $p_C^{*}(I)>p_C(I)$.
In ${\mathcal G}^{*}$ there are no link weights below $I$, therefore,
the list of $k$-cliques with an intensity above $I$ is
identical to the list of all unweighted $k$-cliques.
In other words, the critical point of $k$-clique percolation
in ${\mathcal G}^{*}$ is the same for any value of the intensity threshold
between $0$ and $I$. Specifically, $p_C^{*}(I) = p_C^{*}(0)$.
Moreover, the link deletion step keeps a random $1-I$ portion
of all links from ${\mathcal G}$ and modifies
the unweighted percolation threshold from $p_C(0)$ to
$p_C^{*}(0) = p_C(0)/(1-I)$.
Combining the above gives the following
upper bound for $p_C(I)$:
\bea
p_C(I) < p_C^{*}(I) = \ 
p_C^{*}(0) = \
\f{p_C(0)}{1-I} \, .
\label{eq:up}
\eea

\subsubsection{Branching process,
intensity condition for child $k$-cliques.}

In the second approximation, we treat the percolation of $k$-cliques 
fulfilling the intensity condition as a branching
process visiting $k$-cliques via $k$-clique adjacency connections.
We investigate one branching event:
having arrived at a $k$-clique (parent),
we try to move on to further ones fulfilling $I_{\mathcal C}>I$ as 
 well (children).
Consider one of these child $k$-cliques
and assume that the
probability distribution of each link weight in the parent $k$-clique 
is the original uniform distribution on the interval $(0,1]$.
(The actual probability distribution of a link weight in the parent
$k$-clique is different from this.)

The expected number of all neighbouring $k$-cliques, including those 
with intensities below $I$, is $p^{k-1}N(k-1)$ in the large $N$ limit.
Now apply the intensity condition (Sec.\,\ref{subsec_cpmw})
to each child $k$-clique separately:
we denote by ${\mathcal P}_k ( < 1 )$ the probability that the child $k$-clique
has an intensity larger than $I$. With this notation the expected number of 
accepted  child $k$-cliques available at the current branching step 
is $p^{k-1}N(k-1){\mathcal P}_k$. On the other hand,
being at the critical point means that the expectation value of this number 
should be $1$.
In summary, compared to the $I=0$ (unweighted) case,
we get the following approximation:
\bea
p_C(I) \simeq p_C(0)\,{\mathcal P}_k^{-1/(k-1)} \, ,
\label{eq:lo1}
\eea
\noindent
where ${\mathcal P}_k$ is the 
probability that the product of $k(k-1)/2$ independent
link weights, with uniform distribution on $(0,1]$,
reaches $A = I^{\,k(k-1)/2}$.
For $k=3$ and $4$, the ${\mathcal P}_k$ probabilities are
\bea
{\mathcal P}_3 &=& \ \!\!
\int\limits^{1}_{A}                \!         dw_3 \!\! \
\int\limits^{1}_{\f{A}{w_2}}       \!       dw_2 \!\!\!\!\! \
\int\limits^{1}_{\f{A}{w_3w_2}}   \!\!\!\!   dw_1 = \! \
1 - A\bigg( 1 -\ln A + \f{\ln^2 A}{2}\bigg) \, ,
\nonumber\\
{\mathcal P}_4 &=& \
\int\limits^{1}_{A}                \!       dw_6 \!\! \
\int\limits^{1}_{\f{A}{w_6}}       \!       dw_5 \!\! \
\dots \!\!\!\!\!
\int\limits^{1}_\f{A}{w_6\dots w_2} \!\!\!\!\!\! dw_1  = \
1 - A \sum\limits_{i=0}^{5} \f{(-\ln A)^i}{i!} \, .
\label{eq:lo2}
\eea
\noindent
In summary,
the transition point, $p_C(I)$,
can be approximated in the $k=3$ and $4$ cases
(with $n=k(k-1)/2$) as
\bea
\f{p_C(I)}{p_C(0)}\bigg|_{k=3,4} \simeq \bigg[ 1 - I^{\,n} \
\sum\limits_{i=0}^{n-1} \f{(-n\ln I)^i}{i!} \
\bigg]^{-1/(k-1)} \, .
\eea

\subsubsection{Branching process,
child and first parent $k$-cliques.}
\label{subsec_branching2}

\begin{figure}[t!]
\centerline{\includegraphics[angle=0,width=0.68\columnwidth]{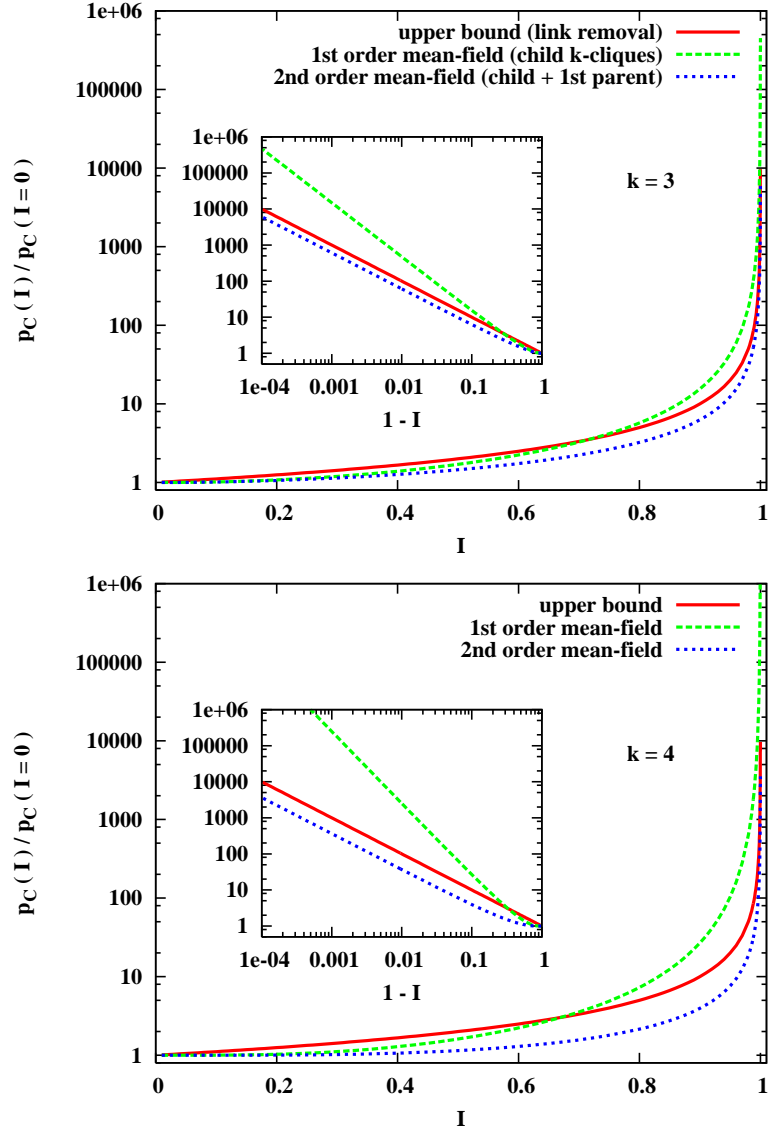}}
\caption[]{
{\bf Main panels.}
Analytical approximations for the critical link probability,
$p_C(I)$, of $k$-clique intensity percolation
in weighted Erd\H os-R\'enyi (ER) graphs 
as a function of the 
intensity threshold, $I$
(see text for details).
Clique size parameters are $k=3$ (top) and $k=4$ (bottom).
We plotted the ratio between $p_C(I)$ and the critical link
probability, $p_C(0)$, of clique percolation without weights
\cite{derenyi05prl}.
In the ER graph 
each link is filled with probability $p$ and link
weights are randomly and uniformly
selected from the interval $(0,1]$.
{\bf Insets.}
The same curves transformed.
At low $I$
the first order (dashed green) and second order mean-field
(dotted blue) approximations 
are below the upper bound (solid red),
while for $I\to 1$,
the first order approximation diverges faster than
the strict upper bound. 
We suggest that for each $k$ increasing
the precision of the approximations in Sec.\,\ref{subsec_an}
(to $3$rd, $4$th, \etc\,order)
will make the solution converge to the exact one.
We predict that for the
exact solution $p_C(I)/p_C(0)$ 
diverges as $(1-I)^{-1}$ when $I\to 1$.
}
\label{fig_ERanalytical}
\end{figure}

We improve the previous approximation and modify
${\mathcal P}_k$ by taking into account that the parent
$k$-clique has an intensity above $I$.
Due to this condition the 
the distributions of the $(k-1)(k-2)/2$ link weights
in the overlap
(connecting the $(k-1)$ shared nodes of the parent $k$-clique
and its child) are not independent from each other.
The distribution density of the product, $t$,
of these link weights is
\bea
\tilde{p}_k(t) = \f{f_k(t)}{C} = \
\f{1}{C} \int\limits_{A/t}^{1}\!\!dw_1\!\!\!\!\
\int\limits_{A/(tw_1)}^{1}\!\!\!\!\!\!dw_2\
\dots\!\!\!\!\!\!\!\!\!\!\!\!\
\int\limits_{A/(tw_1\dots w_{k-2})}^{1}\
\!\!\!\!\!\!\!\!\!\!\!\!\!\!\!\!dw_{k-1}\!\!\ \, .
\eea
\noindent
Each of the integrations is an averaging
for one of the $k-1$
links of the parent $k$-clique not
contained by the overlap.
The normalisation constant is $C=\int_{A}^{1}dt\,f_k(t)$.
To compute the probability that the
child $k$-clique's intensity is above $I$, 
the same integrations should be performed for
the $k-1$ links of the child $k$-clique outside the overlap.
Therefore, we get 
\bea
\f{p_C(I)}{p_C(0)} \simeq \
{\mathcal P}_k^{-1/(k-1)} = \ 
\Bigg(\! \
{\int_{A}^{1}dt\,f_k(t) \over \
 \int_{A}^{1}dt\,f_k^2(t)} \
\!\Bigg)^{1/(k-1)} \!\!\!\! \, .
\label{eq:f_ratio}
\eea

Again, as an example, we have performed the integrals and computed
$p_C(I)$ for $k=3$ and $4$:
\bea
f_3(t) &=& \
1 - \f{A}{t} \bigg( 1- \ln\f{A}{t}\bigg) \
\, ,\nonumber\\
f_4(t) &=& \
1 - \f{A}{t} \bigg( 1 - \ln\f{A}{t} + \f{1}{2}\ln^2\f{A}{t} \bigg) = \
= 1 - \f{A}{t}\sum\limits_{i=0}^{2} \
\f{\big(-\ln\f{A}{t}\big)^i}{i!} \, ,
\nonumber
\eea
\noindent
which gives
\bea
\f{p_C(I)}{p_C(0)}\bigg|_{\scriptstyle k=3} \!\! &\simeq& \!\! \
\bigg[ \
\f{ 1 - I^3( 1 - 3 \ln I + \f{9}{2}\ln^2 I)}{\scriptstyle 1 +
I^3 [ 4 - 5I^3 + 6 (1+2I^3) \ln I - 9 (1+I^3) \ln^2 I]} \
\bigg]^{1/2}  
\eea
\noindent 
and
\bea
\f{p_C(I)}{p_C(0)}\bigg|_{\scriptstyle k=4} \!\! &\simeq& \!\! \
\bigg[ \f{F_4(I)}{G_4(I)} \bigg]^{1/3}\, ,
\eea
\noindent
where
\bea
F_4(I) = 
1 - I^6 \bigg[
1 &-& 6\ln I + 18\ln^2 I - 36\ln^3 I
\bigg] \, ,\\
G_4(I) = 
1 + I^6 \bigg[ 18 &-& 19 I^6 + 12(1+9I^6)\ln I - \
 36(1+8I^6)\ln^2\!I + \nonumber\\
&+& 72(1+6I^6)\ln^3\!I - 324I^6\ln^4\!I
\bigg] \, .\nonumber
\eea

\subsection{Numerical results}
\label{subsec_num}

\begin{figure}[t!]
\centerline{\includegraphics[angle=0,width=0.98\columnwidth]{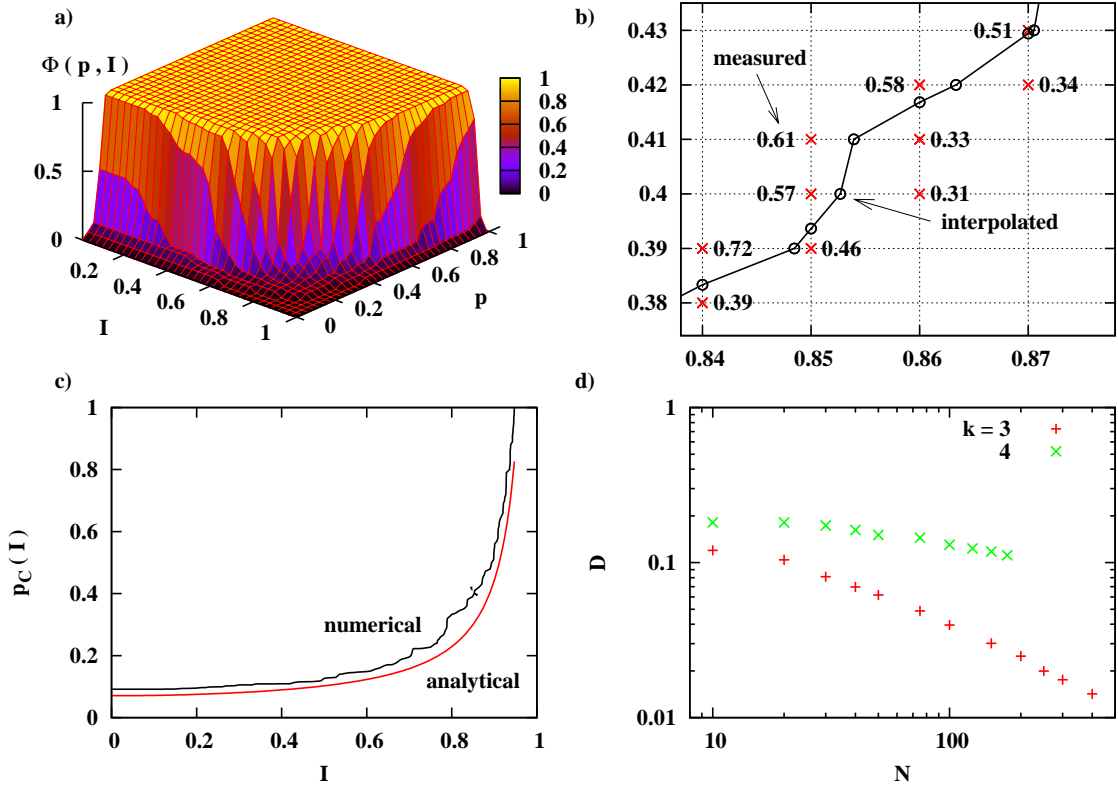}}
\caption[]{
Numerical analysis of the percolation of $k$-cliques fulfilling the
 intensity condition in weighted Erd\H os-R\'enyi graphs.
The sample numerical results shown in panels (a-c) were obtained
for $N=100$ and $k=3$ using $1$ run for each $(p, I)$ grid point.
In panel (d) points were computed
from $3$ to $100$ runs for each $(k, I)$ parameter pair
and error bars are smaller than the sizes of the symbols.
{\bf (a)} 
The order parameter, $\Phi=n_1/\sum_\alpha n_\alpha$,
in the points of a grid on the $(k,I)$ plane.
{\bf (b)}
We computed the transition line,
$p_C=p_C(I)$, as the curve with $\Phi=1/2$ on the $(k,I)$ plane. 
From the values of $\Phi$ at nearby grid points we increased the
precision of the transition line with linear interpolation.
{\bf (c)} 
Numerical curve for the percolation threshold
and the second order analytical approximation
from Sec.\,\ref{subsec_branching2}
The area between the two curves, $D$,
measures the difference between the two results.
{\bf (d)} 
Difference
between the numerical and analytical results for
$p_C(I)$ at various system sizes, $N$,
and clique size parameters. 
}
\label{fig_ERnum}
\end{figure}

We generated weighted ER graphs as described above,
and extracted the $k$-clique percolation clusters emerging from 
 the $k$-cliques fulfilling the intensity condition for 
several threshold ($I$) and clique size parameter ($k$) values.
Denoting again by $n_\alpha$ the number of nodes in a module 
(percolation cluster) $n_1\ge n_2\ge \dots$,
we used as an order parameter the 
relative number of nodes in the largest module:
\bea
\Phi = \f{n_1}{\sum_\alpha n_\alpha} \, .
\eea
\noindent
It is known that in the classical Erd\H os-R\'enyi 
link percolation problem below the critical link probability,
$p_C$, all clusters contain significantly fewer nodes
than the total ($N$), while above $p_C$ there is one module
with size ${\mathcal O}(N)$ and all others are much smaller
\cite{bollobas85book}.
One can measure the transition point between these two regimes in
several ways that are equivalent in the large system size limit.
Here we decided to identify the critical point as the link probability
where the order parameter, $\Phi$, becomes $1/2$.
Fig.\,\ref{fig_ERnum} shows our numerical results for the critical
point of intensity $k$-clique percolation in ER graphs 
and a comparison with the analytical result from
Sec.\,\ref{subsec_branching2}.
To quantify the distance between the numerical
and analytical results, we computed the difference integral, $D$,
between the two curves.
With growing system size $D$ decreases
indicating that the second order approximation
converges to the actual transition curve, $p_C(I)$.

Compared to our generic CPMw search method,
the numerical work presented in this section
was accelerated by a factor of $\approx\!\!100$
with the help of two algorithmic improvements
constructed for this purpose.
We computed the order parameter, $\Phi$,
in all $>1,000$ points of a grid on the $(p, I)$ plane
(Fig.\,\ref{fig_ERnum}a).
Depending on the total number of nodes, $N$,
we used in each grid point
$3$ to $100$ samples (weighted ER networks).

The first algorithmic improvement was based on the observation that
for a fixed graph and a fixed clique size parameter, $k$, 
the weighted modules at two intensity thresholds ($I_1>I_2$)
differ only in the $k$-cliques
with intensities between $I_1$ and $I_2$.
Recall that
the weighted modules at $I_1$ (or $I_2$) contain the $k$-cliques
with intensities above $I_1$ ($I_2$).
Knowing all $k$-cliques with intensities above $I_1$,
one can compute the weighted modules for the threshold $I_2$
by adding $k$-cliques between $I_1$ and $I_2$
and then assembling the percolation clusters of $k$-cliques.
Thus, to find the weighted modules
in a given ER graph at each of the intensity threshold values
$I_1>I_2>\dots>I_n$,
one does not need to perform the entire CPMw and consider all
$k$-cliques again at each $I_i$.
We first listed all $k$-cliques with intensities above $I_n$,
and then sequentially inserted them (into an empty graph)
in the descending order of their intensities.
Whenever we reached an $I_i$ threshold, we assembled the weighted
modules based on those already computed for the previous threshold,
$I_{i-1}$ in an analogous way to the Hoshen-Kopelman algorithm
\cite{hoshen76prb}.
During the process of inserting $k$-cliques
if the size of the largest module reached $N$, \ie,
the order parameter, $\Phi$, became $1$, 
then we set $\Phi=1$ for all lower
$I_i$ thresholds and proceeded to the
next parameter set.

The second algorithmic improvement allowed us to find $k$-clique
adjacencies in shorter time and thereby to assemble the percolation
clusters of $k$-cliques faster.
If a $k$-clique overlaps with another $k$-clique,
then they share one of the $(k-1)$-cliques contained by the first.
Thus, we listed the $(k-1)$-cliques occurring in all
considered $k$-cliques, and for each we listed its containing 
$k$-clique(s).
More than one containing $k$-clique
for a $(k-1)$-clique
means that the containing $k$-cliques are all pairwise adjacent.
Note also that 
all $k$-clique adjacency connections can be located this way.

\section{Results for real-world graphs}
\label{sec_real}

As opposed to the Erd\H os-R\'enyi model, in real-world graphs 
local properties (\eg, node degree, strength and link weight)
are often correlated giving rise to small-, intermediate- and 
large-scale network structures. Below, we analyse 
link weight correlations and the structure of weighted modules in two
types of real webs. The first is a social (scientific co-authorship)
net and the second is a set of two stock correlation graphs.

\subsection{Scientific co-authorship network (SCN)}
\label{subsec_scn}

Social networks were among the first few where the small-world \cite{milgram67}
 and scale-free  \cite{barabasi99science} properties were observed. Since then
 several models have been constructed to describe these and further 
characteristics \cite{watts98nature,barabasi99science} and some of the 
microscopic rules of the models have been verified by direct measurements on 
real graphs \cite{jeong03epl}. Scientific collaboration networks, as webs of 
professional contacts, are usually ``measured'' through lists of joint 
publications. Here we consider the weighted co-authorship network
of researchers appearing on the $50,634$ e-prints of the Los Alamos cond-mat
archive \cite{warner03lht} between April 1992 and February 2004.
In this graph a paper with $r$ authors contributes by $1/(r-1)$ to the weight 
of the link connecting any two of its authors (nodes) and thus, the strength 
of a node is equal to the number of papers of the author. 
In the resulting weighted co-publication graph there are $31,319$ non-isolated 
nodes with $136,065$ links between them;
these nodes have an average degree (collaborator number) 
of $8.69$ and an average strength (paper number) of $4.47$.

\begin{figure}[t!]
\centerline{\includegraphics[angle=0,width=0.98\columnwidth]{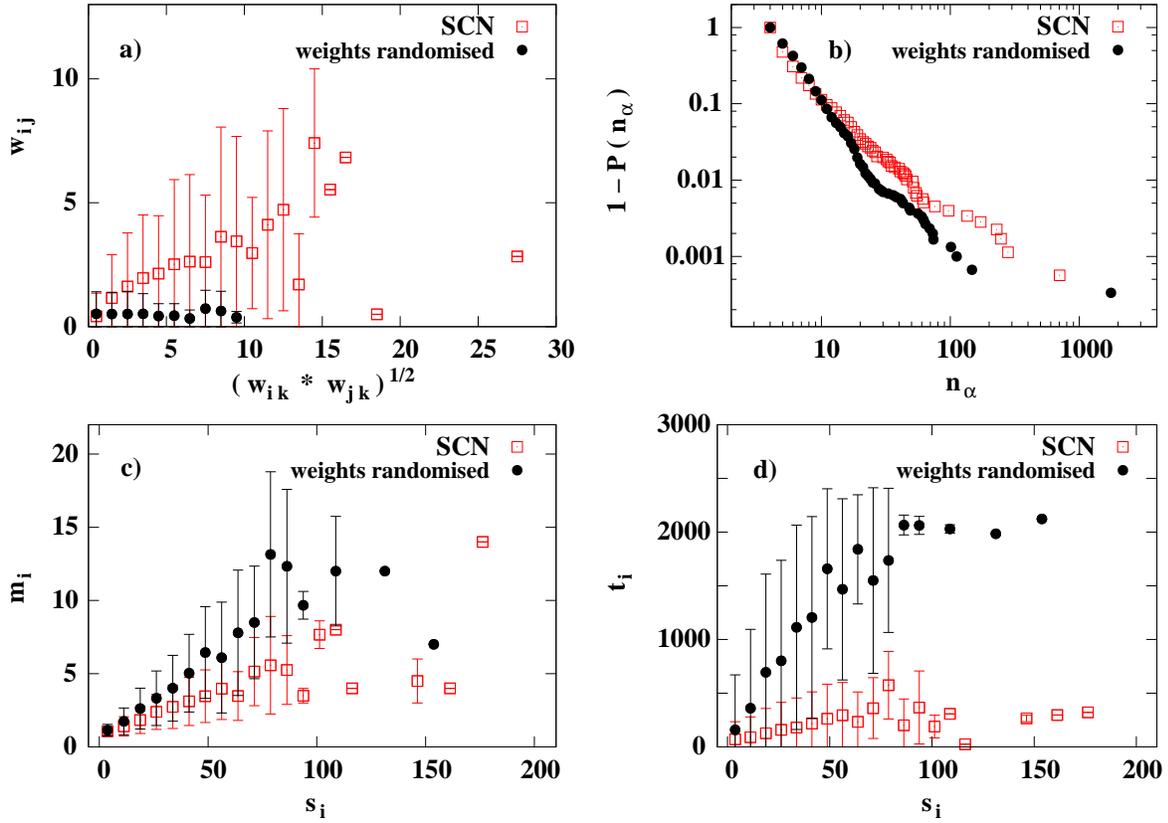}}
\caption[]{
Link weight correlations (in triangles) and weighted modules in
the weighted co-publication network of cond-mat authors.
The randomised control graph was constructed by
shuffling the weights of the links. Different 
instances of the randomised control graph
(with other random seeds) 
produced similar results.
{\bf (a)}
In triangles (nodes $i$, $j$, $k$)
the weight of a link, $w_{ij}$,
grows roughly linearly
with the geometric mean of the other two link weights,
$w_{ik}$ and $w_{jk}$.
{\bf (b)}
Cumulated size distribution of weighted modules.
Observe that the largest weighted module of the randomised graph is
significantly larger than that of the SCN.
{\bf (c)}
Except for scientists with $s_i>80$ publications,
the number of communities (modules) of a node (author)
grows linearly with its strength (paper number),
similarly to {\bf (d)} the number of co-authors, $t_i$,
contained by these communities. 
}
\label{fig_scn}
\end{figure}

Several correlation properties of the SCN (both unweighted and weighted) 
are well-known from previous studies. As for the unweighted case,
node degrees are assortative and the clustering coefficient is high
\cite{newman01pre1}. Moreover, nodes with the highest degrees tend to 
form so-called rich-clubs \cite{colizza06natphys,zhou04ieee},
\ie, they are more likely to be linked to each other than
in the corresponding fully uncorrelated (ER) model. The weighted correlation 
measures of the SCN analysed so far have been $2$- and $3$-point correlation 
functions, which were found to be influenced mainly by the positions of the 
graph's links, but not the weights of the links 
\cite{barrat04pnas,serrano06pre}. The expected weight of a link is almost 
independent from its end point degrees. Weighted nearest-neighbour degree 
correlations and weighted clustering coefficients have highly
similar distributions to the analogous unweighted quantities
both as a function of node degree and strength.
The difference between the investigated weighted and
unweighted measures 
was found to be
much smaller in the SCN than in other types of real webs,
\eg, air transportation and trade networks.

Here we show that there are correlation properties of the SCN significantly 
influenced by the links' weights, not only by the positions of the links.
The information contained by the link weights can be decomposed into two parts.
The first is the (heavy-tailed) distribution of the weights and the second is 
how these numbers are arranged on the links of the underlying unweighted graph.
We constructed a randomised null model, a control graph, of the SCN. We kept 
the positions of links (a list of node pairs) and the list of link weights 
(non-negative numbers) unchanged and shuffled the weights on the links of 
the graph. Comparing the SCN to its control graph, we found a strong 
assortativity of link weights in triangles (Fig.\,\ref{fig_scn}a): two links 
with high weights have a third neighbouring link with a high weight, too.

The tendency of high link weights to stay close to each other can be measured 
for groups containing more than three links as well. A standard tool for 
analysing such correlations is provided by enumerations methods
listing each possible subgraph of a fixed size. Along this approach, we used 
the CPMw to compute the weighted overlapping modules for the SCN and its 
randomised counterpart, and inferred link weight correlation properties 
by comparing the sizes of the obtained modules in the two systems. 
The optimal intensity threshold and $k$-clique size parameters 
for the SCN were found to be $I=0.439$ and $k=4$. The largest weighted module 
contained $n_1^{(SCN)}=714$ authors, whereas in case of the randomised graph
 (at the same $I,k$ parameters) we observed $n_1^{(rnd)}=1,946$ 
(Fig.\,\ref{fig_scn}b).
The $n_1^{(SCN)}<n_1^{(rnd)}$ relation indicates that large link weights
 cluster together more strongly in the largest component of the SCN
 than expected by chance: the more closely large ($w_{ij}>I$) link weights 
cluster together, the smaller the number of $k$-cliques will fulfil the 
intensity condition  and the smaller the largest weighted module becomes.
For comparison, we computed the modules of the original CPM in the
 SCN as well, at the same $k$-clique size ($k=4$) and a link-weight 
threshold $W=I$.
About 32\% of the CPM communities were exactly the same in the CPMw approach,
 and a further 27\% were contained in a larger CPMw module.

The CPMw allows overlaps between the modules which enables the investigation 
of further weighted correlation properties. In Figs.\,\ref{fig_scn}c-d we 
quantify the influence of  strong hubs (researchers with many publications)
on the densely internally coupled 
modules of their co-authors. We find that except for authors with very 
large paper numbers both the number of communities, $m_i$, and the number 
of module neighbours, $t_i$, of a scientist grow roughly linearly with the 
number of his/her  publications. (Note that $t_i$ is the number of co-authors
 in dense communities, which is usually smaller than the total number of 
co-authors, $d_i$, \ie, the degree of the node.) However, both $m_i$ and $t_i$ 
remain well below the values obtained for the randomised case.
These findings indicate that authors remain focussed over time
and maintain tight collaborations only with a relatively small number of 
colleague groups. This weighted correlation behaviour can be quantified more 
accurately with intermediate-scale methods, \eg, weighted module finding 
algorithms, than previous $2$- or $3$-node weighted correlation measurements.
Figure\,\ref{fig_scn}c shows that among authors with $s_i>80$ publications 
the average number of modules of one author is above $4$.

\subsection{Correlation graphs of NYSE stocks}
\label{subsec_stock}

\begin{figure}[t!]
\centerline{\includegraphics[angle=0,width=0.98\columnwidth]{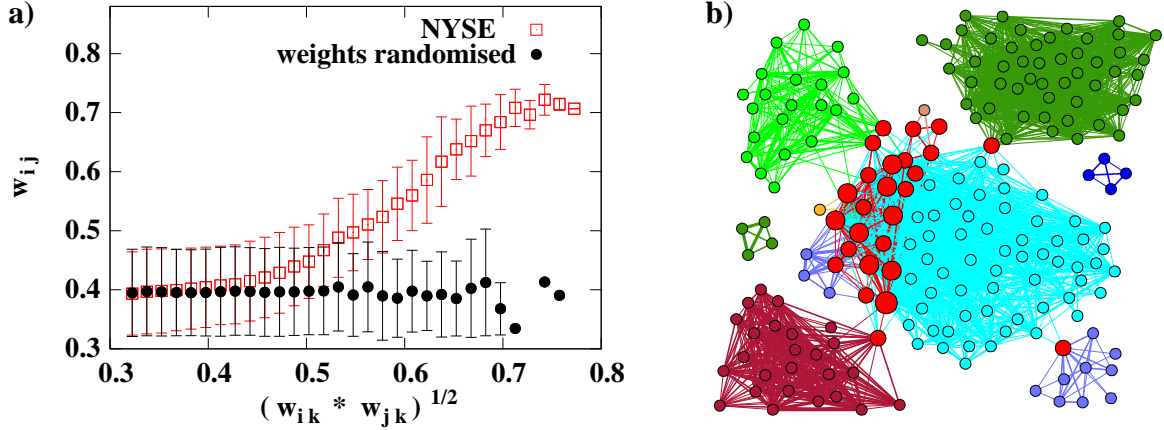}}
\caption[]{
{\bf (a)}
In the NYSE stock graph
two strong links of a triangle have a strong third neighbour.
{\bf (b)}
Weighted modules of the stock graph.
Each node is coloured according to its module.
A node contained by more than one module is coloured red
and its size is proportional to the number of modules
it is contained by.}
\label{fig_stock}
\end{figure}

Financial markets, similarly to the participants of a social web,
integrate information from a multitude of sources and are truly
complex systems. 
The most widely investigated subunits 
of a market are its individual stocks ($i$)
and their performances are measured by their prices,
$P_i(t)$, over time.
Common economic factors influencing the prices of two selected stocks
(nodes) are usually detected from the (absolute) value of their
correlation (weighted link),
which allow one to assemble a network of stocks.
In the statistical physics literature minimum spanning trees
and asset graphs defined on this web have been have been applied to
uncover the hierarchical structure of markets 
\cite{mantegna99epjb}
and their clustering properties
\cite{onnela04epjb}.
Notably, the correlations in their original, matrix, form also provide
useful insights when compared to random matrix ensembles as controls
\cite{laloux99prl,plerou99prl}.

We have analysed a pre-computed
stock correlation matrix \cite{onnela05pre}
containing averaged correlations between the daily logarithmic returns,
$r_i(t)=\ln P_i(t)- \ln P_i(t-1)$, of $N=477$ NYSE stocks.
Considering a time window of length $T$,
one can compute the equal time correlation coefficients
between assets $i$ and $j$ as
\bea
c_{ij}(t)=\
\f{\langle r_i(t)r_j(t)\rangle-\
\langle r_i(t)\rangle\langle r_j(t)\rangle}{\
[\langle r_i^2(t)\rangle-\langle r_i(t)\rangle^2  \big]^{1/2}\
\big[\langle r_j^2(t)\rangle-\langle r_j(t)\rangle^2  \big]^{1/2}} \, .
\eea
\noindent
The pre-computed matrix contained the time averages,
$c_{ij}$, of the correlation coefficients over a four-year
period, $1996$ to $2000$ ($T=1,000$ days).
We used each correlation coefficient, $c_{ij}$, as a link weight
between nodes $i$ and $j$. 
As observed and analysed in detail previously in, \eg,
Ref.\,\cite{onnela04epjb}, 
only the strongest links (correlations) convey significant
information, thus, in both cases
we kept only the strongest $3\%$ of all link weights.
The resulting network
had $301$ nodes and $3,405$ weighted links,
the highest and lowest link weights
were $0.786$ and $0.321$. 

Similarly to the previous section,
we constructed a randomised control graph
by reshuffling link weights
to analyse weight correlations
in groups of three and more weights
(Fig.\,\ref{fig_stock}).
We found that in triangles 
the presence of two strong links implies that the third link is also
strong, \ie, groups of $3$ strong links prefer to cluster together.
We computed the weighted modules of the stock graph and 
its randomised control with the CPMw using 
the same $(k,I)$ parameters
and found that the largest modules contained
$s_1^{(NYSE)}=84$ and $s_1^{(rnd)}=190$ nodes,
\ie, the largest module is bigger in the randomised 
control graph than in the original one.
Following the reasoning in Sec.\,\ref{subsec_scn},
this indicates that groups of $2$, $3$
and more strong links prefer to
cluster together in the stock correlation network.

\section{Conclusions}

We have introduced a module identification technique for weighted networks
based on $k$-cliques having a subgraph intensity higher 
than a certain threshold, and allowing shared nodes (overlaps) between modules.
With this algorithm, the CPMw, we first considered 
the percolation of $k$-cliques fulfilling the intensity condition on 
(weighted) Erd\H os-R\'enyi graphs. For the critical link probability 
we showed analytical approximations together with detailed numerical results
and found a quickly decaying difference between the two with growing system 
size.

For two weighted real-world graphs we analysed link weight correlations
within groups of $3$ and more links. The first was a scientific co-authorship 
network (SCN) and the second was a stock correlation graph (NYSE).
In the SCN the weighted $2$ and $3$-point correlation functions studied 
earlier showed only minor differences from the analogous unweighted 
correlation functions. Here we investigated the correlations of weights 
in triangles and computed the weighted modules of the empirical graphs
(SCN and NYSE) with the CPMw.
We found that in both graphs groups of $3$ and 
more strong links cluster together, \ie,
the weighted correlation functions
of $3$ or more links significantly differ from their
randomised counterparts.

\section{Acknowledgements}

We thank I. Der\'enyi for helpful suggestions 
and critical reading of the analytical results.
We thank S. Warner for the ArXiv preprint listings
and J.-P. Onnela and J. Kert\'esz 
for the stock market data.
We acknowledge financial support from
the Hungarian Scientific Research Fund
(grants No. K068669, PD048422 and T049674).

\end{document}